\documentclass[epjCONF,columns]{svjour}
\usepackage{graphics}
\usepackage[varg]{txfonts} % Times fonts
\usepackage[utf8]{inputenc}
\usepackage[colorlinks]{hyperref}
\usepackage{atlasphysics}
\usepackage{lineno}
\session-title{Hadron Collider Physics symposium (HCP 2011)}
\begin{document}
%\linenumbers
%
\title{Di-boson results at ATLAS}
\author{Pierre-François Giraud\thanks{\email{pierre-francois.giraud@cea.fr}} On behalf of the ATLAS Collaboration }
\institute{DSM/IRFU (Institut de Recherches sur les Lois Fondamentales de
l’Univers), CEA Saclay (Commissariat à l’Energie Atomique), Gif-sur-Yvette,
France}
\abstract{ Pairs of gauge boson produced in proton-proton collisions at a
center-of-mass energy $\sqrt{s}$ of $7~\tev$ are reconstructed with the ATLAS
detector in their leptonic final states. Based on samples of integrated
luminosity $\mathcal{L}=1.0~\ifb$ (for $WW$, $WZ$ and $ZZ$) and $35~\ipb$ (for
$W\gamma$ and $Z\gamma$) of 2011 and 2010 LHC data, the total di-boson
production cross sections are measured. They are found, together with the
kinematic distributions of the selected di-boson systems to be compatible with
the expectation from the Standard Model. The di-boson production also gives a
handle on possible anomalous triple gauge boson couplings, for which 95\%
confidence limits are set. }
\maketitle
\section{Introduction}
\label{intro}
In the Standard Model (SM), the triple gauge boson couplings (TGCs) are fully
constraint by the electroweak symmetry. In particular, the $ZZZ$, $ZZ\gamma$
and $Z\gamma\gamma$ neutral TGC vertices are are absent, whereas the $WWZ$ and
$WW\gamma$ vertices are predicted non-zero. For this reason, the measurement of
the di-boson final states at the LHC provides an important test of the SM:
beyond Standard Model physics could contribute to the TGCs and result in
modified di-boson cross sections or final state kinematics. Furthermore,
non-resonant di-boson productions are a background to the search for the Higgs
boson, so it is essential to understand their detection sensitivity.

This note presents measurements of the di-boson production in proton-proton
collisions at a center-of-mass energy $\sqrt{s}$ of $7~\tev$, with the ATLAS
experiment~\cite{RefAtlasExperiment}: their production cross sections are
measured, and first limits on anomalous triple gauge boson couplings (aTGCs) are
set. A sample of integrated luminosity $\mathcal{L}=1.0~\ifb$ of 2011 LHC data
was used to measure the $ZZ$, $WZ$ and $WW$ final states, and
$\mathcal{L}=35~\ipb$ of 2010 data for the $Z\gamma$ and $W\gamma$ final states.
Presently, the only decay modes used to reconstruct these final states are $Z\to
ll$ and $W\to l\nu$ (with $l=e$ or $\mu$): the branching fractions are small,
but the experimental signatures are clean.

\section{Electrons, muons and photons}
One of the important categories of backgrounds of the analyses presented in this
note is arising from the QCD processes: a jet may produce a fake prompt lepton or
photon signal. For example, pions may be mis-identified as electrons or photons.
Another example is heavy flavour jets, which may result in real leptons in the
final state. The main tools to reject this background are cuts on the lepton and
photon identification quantities provided by the detector, which may be
tightened if necessary, and cuts on the isolation energy (the sum of the
transverse energies measured by the calorimeter or the inner detector, in a cone
of fixed size around the candidate lepton or photon).

The probability that particles from jets pass the lepton and photon
identification and isolation cuts is so small that it would be both impractical
and inaccurate to estimate this background with Monte-Carlo (MC) simulation.
Instead, the analyses presented in this note rely on a data-driven method.

A control region enriched in events from the QCD process is built using the
full selection of the chosen final state, except that the isolation or
identification cuts are reversed. The event yield observed in the control
region is extrapolated to the signal region, by the use of a fake factor. The
fake factor needs to be estimated in an independent QCD control sample: for
example a sample of di-jet triggered events, or, if available, a sample
obtained by reversing another of the analysis cuts.

The estimation of this fake factor is in general a significant source of
systematic error: in particular the fake factor varies with data taking
conditions (instantaneous luminosity and pile-up), and the control region may
inaccurately describe the jet content of the signal region (heavy to light jet
ratio for example).

\section{$WW\to l\nu l\nu$}
The $WW\to l\nu l\nu$ signal is measured in final states with two leptons and
missing transverse energy (\met)~\cite{RefWW}. Background discrimination and
estimation is challenging as several processes may fake this final state,
either because they contain real leptons and \met (top events, discriminated by
a jet veto), or fake \met (Drell-Yan, discriminated with dilepton mass veto and
tight \met cut) or jets misidentified as prompt leptons ($W+\mathrm{jet}$
events, discriminated with tight lepton identification and isolation cuts).
After applying a tight selection, $414$ events are observed in a sample of
integrated luminosity $\mathcal{L}=1.0~\ifb$, for a total estimated background
of $169.8 \pm 6.4 \mathrm{(stat.)} \pm 27.1 \mathrm{(syst.)}$ events,
determined with a combination of data-driven and MC techniques. The dominant
systematics are coming from the background estimation: the uncertainty arising
from the jet veto, and from the fake prompt lepton estimation.

\begin{figure}
  \centering
  \resizebox{\columnwidth}{!}{\includegraphics{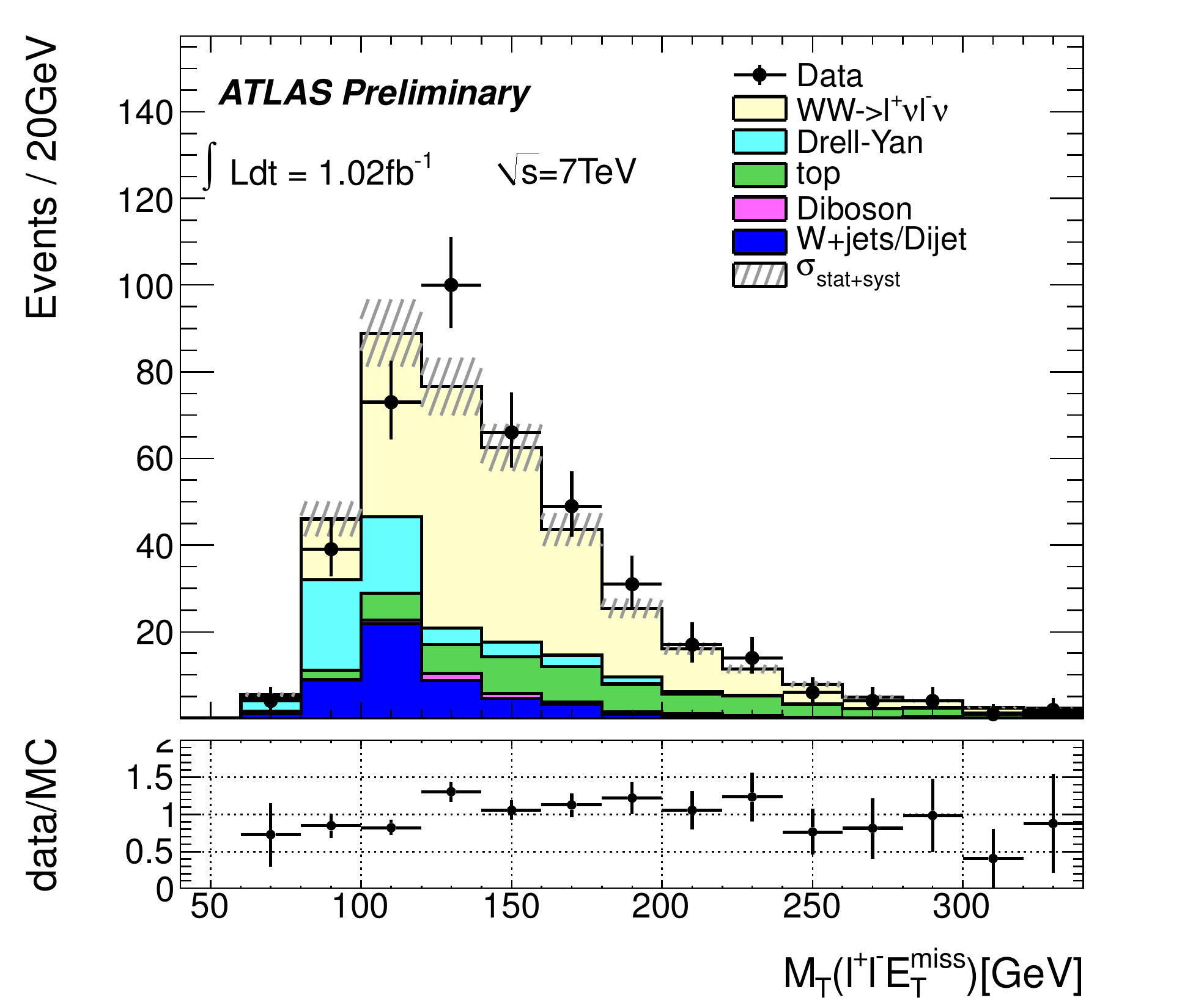}}
  \caption{Distribution of the transverse mass of the di-lepton plus missing
  transverse energy system for $WW$ candidates. The points are the data and the
  stacked histograms are from MC predictions except the $W$+jets background, which
  is obtained from data-driven methods. The estimated uncertainties are shown as
  the hatched bands.}
  \label{fig:ww_mT}
\end{figure}

Figure~\ref{fig:ww_mT} presents the distribution of the transverse mass of the
di-lepton plus \met system after selection: no significant deviation from the SM
expectation is observed. The event yield is converted to the total
cross section $\sigma(pp\to WW) = 48.2 \pm 4.0 \mathrm{(stat.)} \pm 6.4
\mathrm{(syst.)} \pm 1.8 \mathrm{(lumi.)}~\mathrm{pb}$, which is compatible with
the SM prediction at next-to-leading order (NLO) of $46 \pm 3~\mathrm{pb}$.

\section{$WZ \to l\nu ll$}
$WZ$ candidates are selected from events containing three isolated leptons ($e$
or $\mu$) and missing transverse energy~\cite{RefWZ}, where two of the leptons
have a corresponding invariant mass compatible with an on-shell $Z$ decay.  In
comparison with the $WW$ analysis, the $WZ$ analysis benefits from a
three-lepton requirement which reduces most of the background. For that reason,
and in order to increase the analysis acceptance, slightly looser lepton
requirements are applied for the leptons making the Z candidate than for the one
entering the $W$. After selection, a total of $71$ events are observed in a
sample of integrated luminosity $\mathcal{L}=1.0~\ifb$.  The total estimated
background is of $10.5^{+3.0}_{-2.2}$ events, composed of $Z+\mathrm{jet}$ and
top events with a jet mis-identified as a lepton (estimated with a data-driven
technique), and $ZZ\to llll$ (estimated from the MC).

\begin{figure}
  \centering
  \resizebox{0.8\columnwidth}{!}{\includegraphics{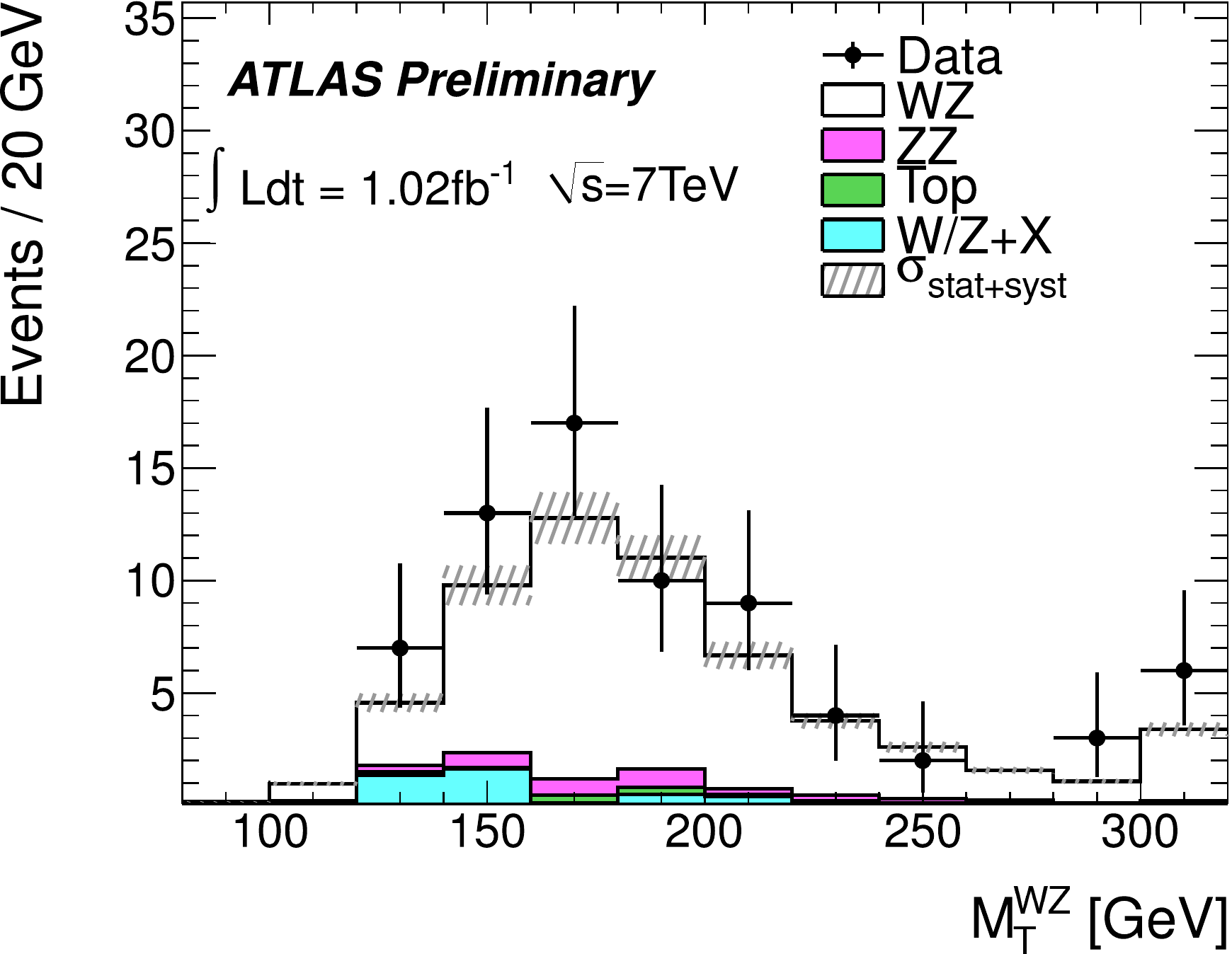}}
  \caption{Distribution of the transverse mass of the $WZ$ system after all cuts
  have been applied. The points represent observed event counts with statistical
  errors, whereas the stacked histograms are the predictions from simulation
  including the statistical and systematic uncertainty. The last bin is an
  overflow bin.}
  \label{fig:wz_mT}
\end{figure}

Figure~\ref{fig:wz_mT} shows the distribution of the transverse mass of the three
leptons plus \met system after selection. The distribution is compatible with
the expectation from the SM. The event yield is converted to the cross section
$\sigma(pp\to WZ) = 21.1 {}^{+3.1}_{-2.8} \mathrm{(stat.)} \pm 1.2
\mathrm{(syst.)} {}^{+0.9}_{-0.8} \mathrm{(lumi.)}~\mathrm{pb}$, compatible
with SM prediction at NLO of $17.2^{+1.2}_{-0.8}~\mathrm{pb}$.

\begin{figure}
  \centering
  \resizebox{0.8\columnwidth}{!}{\includegraphics{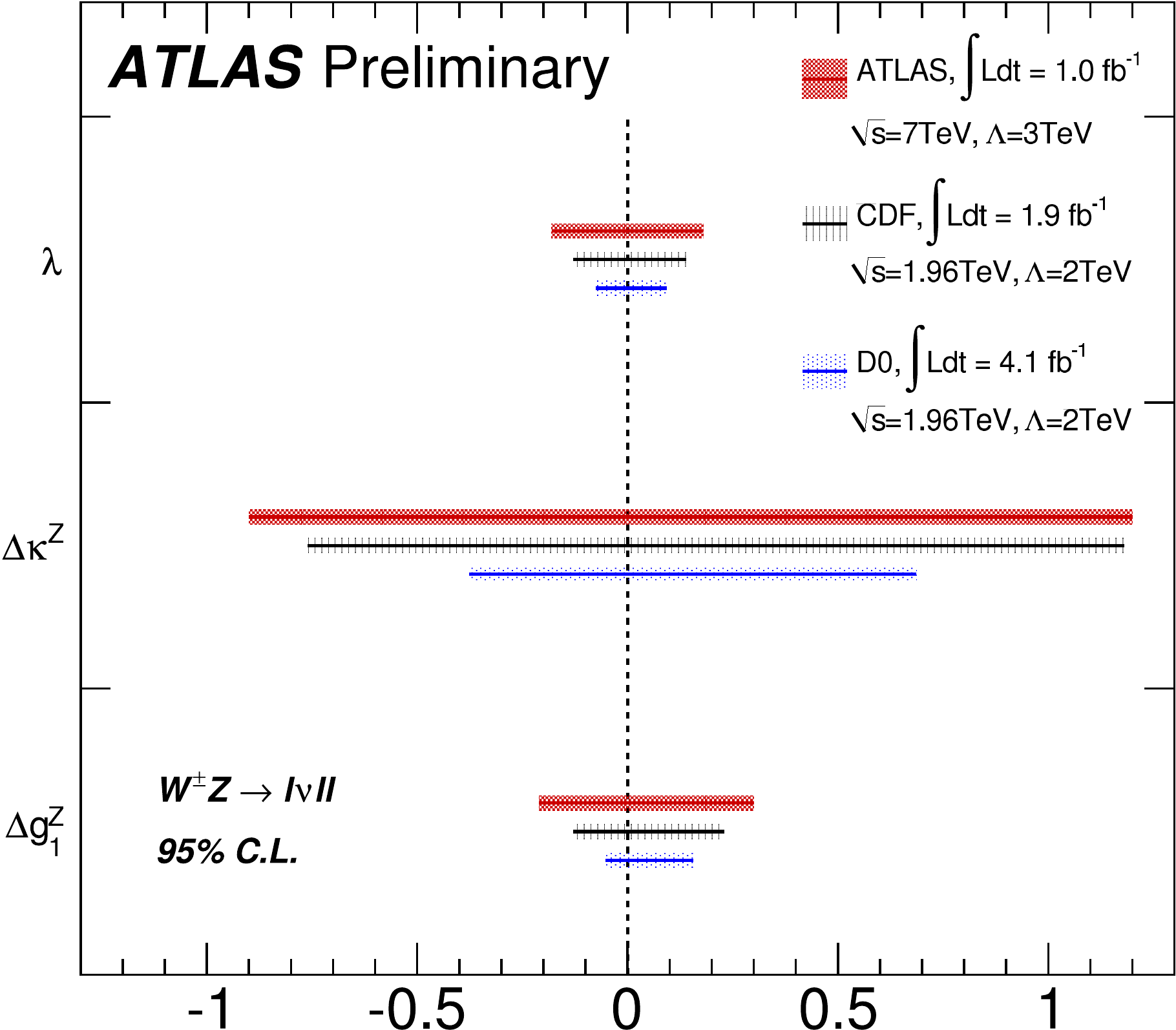}}
  \caption{Limits on aTGC from ATLAS and Tevatron experiments.
  CDF~\cite{RefWZCDF} and D0~\cite{RefWZD0} limits are for $WZ$ production with
  a $\pT(Z)$ shape fit; ATLAS limits are for a cross section fit. Luminosities,
  centre-of-mass energy and cut-off $\Lambda$ for each experiment are shown and
  the limits are for 95\% C.I. }
  \label{fig:wz_tgc}
\end{figure}

$WZ$ events are additionally used to research for possible aTGC terms.
Expressions for the most general effective Lagrangian for a TGC vertex may be
found in~\cite{RefHagiwara} and~\cite{RefEllison}. Three terms of this effective
lagrangian describing aTGCs are presently accessible with the ATLAS $WZ$ data:
$\lambda$, $\Delta \kappa^Z$ and $\Delta g_1^Z$. The $WZ$ cross-section
measurement is used to determine 95\% frequentist confidence intervals on these
three terms, which are shown in figure~\ref{fig:wz_tgc}, and are compared to
Tevatron limits. The limits set by ATLAS are compatible with those of Tevatron,
which have the best sensitivity at the moment. In the future, ATLAS will use the
information in the kinematic distributions of the $WZ$ system to improve these
limits.

\section{$ZZ\to llll$}
The researched signature for the $ZZ$ signal~\cite{RefZZ} is events containing two
pairs of isolated leptons ($e$ or $\mu$), compatible with on-shell $Z$ decays.
Events with four leptons at the LHC are extremely rare, making the $ZZ$ analysis
effectively background free. A total of twelve events are observed for an integrated
luminosity $\mathcal{L}=1.0~\ifb$. The backgrounds considered in this analysis
are inclusively events $lllj$ or $lljj$, where $j$ is a jet mis-identified as a
lepton (such final states may be found in top and $Z+\mathrm{jet}$ events). Most
of these events are rejected by the isolation requirement. A data-driven
technique is used to estimate this background and results in $0.3 \pm 0.3
\mathrm{(stat.)} {}^{+0.4}_{-0.3} \mathrm{(syst.)}$ events.

\begin{figure}
  \centering
  \resizebox{0.9\columnwidth}{!}{\includegraphics{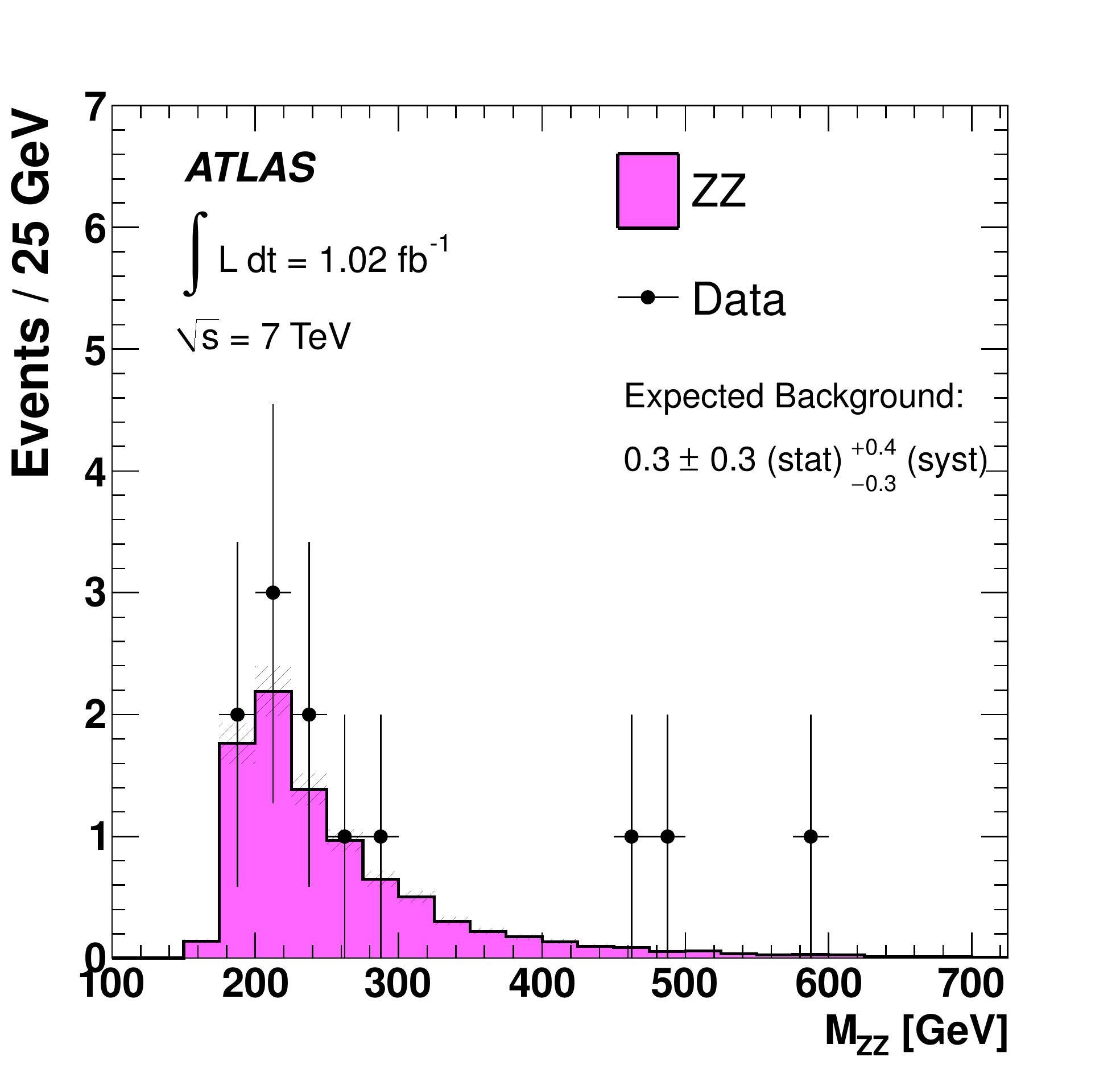}}
  \caption{Invariant mass of the four-lepton system for the selected $ZZ$
  events. The points represent the observed data and the histograms show the
  signal prediction from simulation. The shaded band on each histogram shows the
  combined statistical and systematic uncertainty on the signal prediction. The
  predicted number of background events from the data-driven background estimate
  is indicated on the plot. }
  \label{fig:zz_m}
\end{figure}

Figure~\ref{fig:zz_m} presents the invariant mass distribution of the four-lepton
system after selection. Although three events are observed at an invariant mass
$M_{ZZ}>400~\gev$, the distribution is estimated to be compatible with the SM
expectation. The event yield is used to estimate the total cross section
$\sigma(pp\to ZZ) = 8.5 {}^{+2.7}_{-2.3} \mathrm{(stat.)} {}^{+0.4}_{-0.3}
\mathrm{(syst.)} \pm 0.3 \mathrm{(lumi.)}~\mathrm{pb}$, consistent with the SM
prediction at NLO of $6.5^{+0.3}_{-0.2}~\mathrm{pb}$.

\begin{figure}
  \centering
  \resizebox{0.9\columnwidth}{!}{\includegraphics{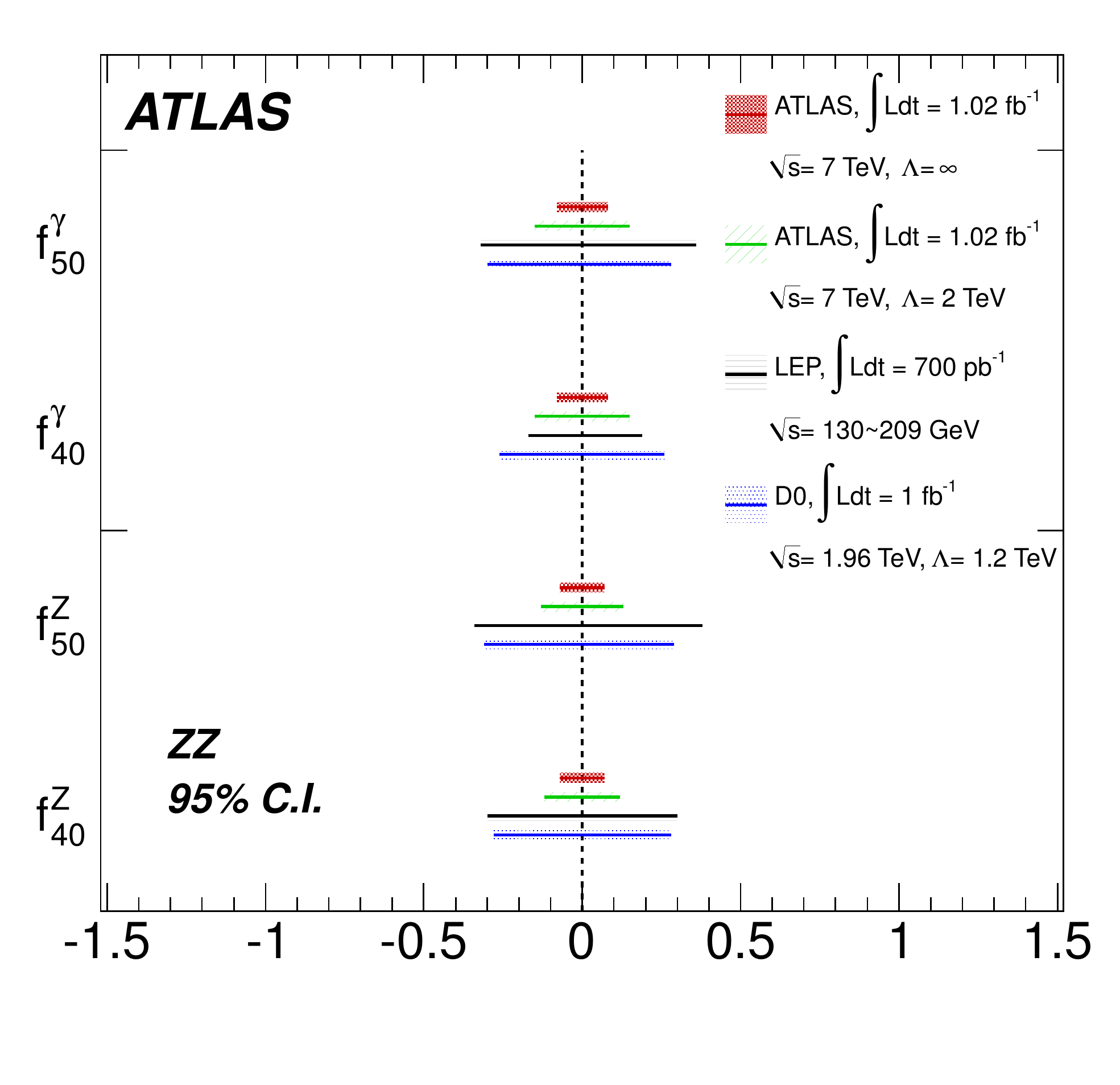}}
  \caption{Anomalous neutral TGC 95\% confidence intervals from ATLAS,
  LEP~\cite{RefZZLEP} and Tevatron~\cite{RefZZD0} experiments. Luminosities,
  centre-of-mass energy and cut-off $\Lambda$ for each experiment are shown.}
  \label{fig:zz_tgc}
\end{figure}

Similarly to the $WZ$ analysis, possible aTGC terms are researched using the
selected $ZZ$ event yield. In the general effective lagrangian, four aTGC
vertices are accessible in ATLAS $ZZ$ data: $f_{40}^Z$, $f_{50}^Z$,
$f_{40}^\gamma$ and $f_{50}^\gamma$. 95\% frequentist confidence intervals are
set on these four terms and are presented in figure~\ref{fig:zz_tgc}. The
limits set by ATLAS for neutral TGC terms are compatible, and exceed in
precision, the limits set by previous experiments from LEP and Tevatron.

\section{$W\gamma \to l\nu\gamma$ and $Z\gamma \to ll\gamma$}
The researched signals $W\gamma$ and $Z\gamma$ contain a leptonic $W$ or $Z$
candidate, and a highly energetic photon~\cite{RefVg}. The signal is defined
with phase space cuts on the photon energy ($\et^\gamma>15~\gev$), separation
from closest lepton ($\Delta R > 0.7$), and isolation at parton level
($\sum_{\mathrm{parton}} \et(\Delta R<0.4) / \et^\gamma < 0.5$). With these
cuts, 8\% of the signal comes from photons originating from the fragmentation
process, and no effort is made to disentangle them from the photon production of
the hard process.

An integrated luminosity of $\mathcal{L}=35~\ipb$ of 2010 LHC data is analysed
and results in 192 $W\gamma$ candidates and 48 $Z\gamma$ candidates. The
background is estimated with a data-driven technique and accounts for $\sim
29\%$ of the $W\gamma$ selection and $\sim 15\%$ of the $Z\gamma$ selection.

\begin{figure}
  \centering
  \resizebox{\columnwidth}{!}{\includegraphics{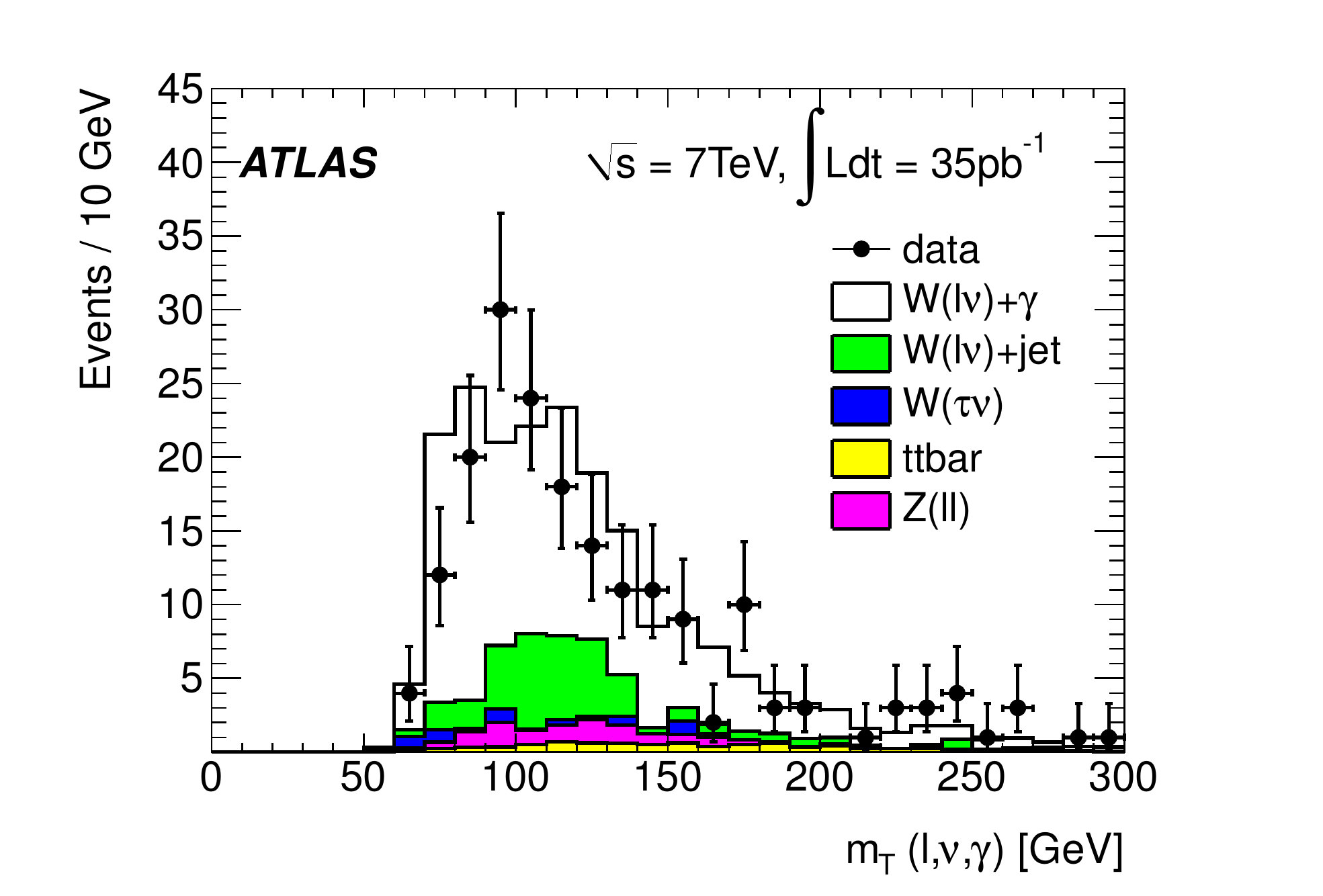}}
  \caption{Distributions for the combined electron and muon decay channels of
  the three body transverse mass ($m_{\mathrm{T}}(l,\nu,\gamma)$) of the $W\gamma$
  candidate events. MC predictions for signal and backgrounds are also shown.}
  \label{fig:wg_mT}
\end{figure}
\begin{figure}
  \centering
  \resizebox{\columnwidth}{!}{\includegraphics{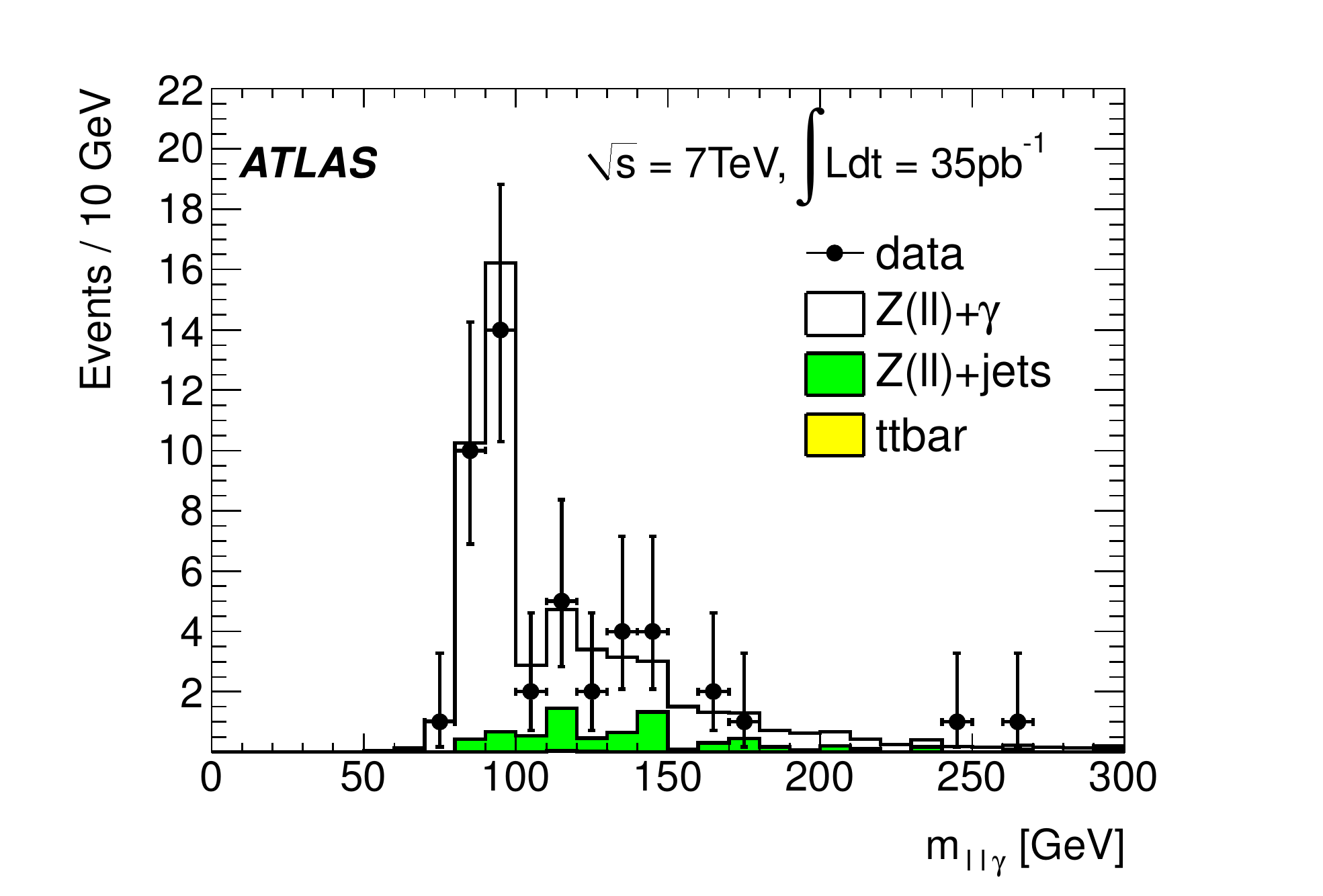}}
  \caption{Three body invariant mass $m(l^+,l^-,\gamma)$ distribution for
  $Z\gamma$ data candidate events. MC predictions for signal and backgrounds are
  also shown. Both the electron and muon decay channels are included.}
  \label{fig:zg_m}
\end{figure}

Figure~\ref{fig:wg_mT} presents the distribution of the transverse mass of the
$l\nu\gamma$ system for the selected $W\gamma$ candidates, and
figure~\ref{fig:zg_m} the invariant mass of the $ll\gamma$ system for the
$Z\gamma$ candidates. The kinematic distributions are compatible with the
expectation from the SM. The event yields are converted to cross section
measurements after background subtraction. For $W\gamma$, the cross section is
measured $\sigma(pp\to W\gamma \to l\nu\gamma) = 36.0 \pm 3.6 \mathrm{(stat.)}
\pm 6.2 \mathrm{(syst.)} \pm 1.2 \mathrm{(lumi.)}~\mathrm{pb}$, for a SM
prediction at NLO of $36.0 \pm 2.3~\mathrm{pb}$. For $Z\gamma$, the measurement
is $\sigma(pp\to Z\gamma \to ll\gamma) = 6.5 \pm 1.2 \mathrm{(stat.)}
\pm 1.7 \mathrm{(syst.)} \pm 0.2 \mathrm{(lumi.)}~\mathrm{pb}$, for a SM
prediction of $6.9\pm 0.5$. Both cross sections are compatible with the SM
expectation.

\section{Conclusion}
Measurements of the production cross sections $pp\to WW$, $pp\to WZ$, $pp\to
ZZ$, $pp\to W\gamma$ and $pp\to Z\gamma$ have been performed with the ATLAS
detector at $\sqrt{s} = 7~\tev$ center-of-mass energy, using samples of
$1.0~\ifb$ and $35~\ipb$ of 2011 and 2010 LHC data. The total production cross
sections are compatible with the SM predictions, and the kinematic distributions
of the various di-boson systems do not show evidence of new physics. Since the
di-boson production is sensitive to the predicted three-boson coupling of the
Standard Model, two of the di-boson measurements ($WZ$ and $ZZ$) have been used to
set first ATLAS limits on possible anomalous TGC terms.

\end{document}

% end of file template.tex